\documentclass[RNAAS]{aastex63}

\graphicspath{{./}{figures/}}
\begin{document}

\title{Detection of a glitch in PSR J0908$-$4913 by UTMOST}

\correspondingauthor{Marcus E. Lower}
\email{mlower@swin.edu.au}

\author[0000-0001-9208-0009]{Marcus E. Lower}
\affiliation{Centre for Astrophysics and Supercomputing, Swinburne University of Technology, PO Box 218, Hawthorn, VIC 3122, Australia}
\affiliation{CSIRO Astronomy and Space Science, Australia Telescope National Facility, Epping, NSW 1710, Australia}

\author[0000-0003-3294-3081]{Matthew Bailes}
\affiliation{Centre for Astrophysics and Supercomputing, Swinburne University of Technology, PO Box 218, Hawthorn, VIC 3122, Australia}
\affiliation{OzGrav: The ARC Centre of Excellence for Gravitational-wave Discovery, Hawthorn VIC 3122, Australia}

\author[0000-0002-7285-6348]{Ryan M. Shannon}
\affiliation{Centre for Astrophysics and Supercomputing, Swinburne University of Technology, PO Box 218, Hawthorn, VIC 3122, Australia}
\affiliation{OzGrav: The ARC Centre of Excellence for Gravitational-wave Discovery, Hawthorn VIC 3122, Australia}

\author[0000-0002-7122-4963]{Simon Johnston}
\affiliation{CSIRO Astronomy and Space Science, Australia Telescope National Facility, Epping, NSW 1710, Australia}

\author[0000-0003-1110-0712]{Chris Flynn}
\affiliation{Centre for Astrophysics and Supercomputing, Swinburne University of Technology, PO Box 218, Hawthorn, VIC 3122, Australia}
\affiliation{ARC Centre of Excellence for All-sky Astrophysics (CAASTRO)}

\author{Timothy Bateman}
\affiliation{Sydney Institute for Astronomy (SIfA), School of Physics, The University of Sydney, NSW 2006, Australia}

\author{Duncan Campbell-Wilson}
\affiliation{Sydney Institute for Astronomy (SIfA), School of Physics, The University of Sydney, NSW 2006, Australia}

\author[0000-0002-8101-3027]{Cherie K. Day}
\affiliation{Centre for Astrophysics and Supercomputing, Swinburne University of Technology, PO Box 218, Hawthorn, VIC 3122, Australia}
\affiliation{CSIRO Astronomy and Space Science, Australia Telescope National Facility, Epping, NSW 1710, Australia}

\author[0000-0001-9434-3837]{Adam Deller}
\affiliation{Centre for Astrophysics and Supercomputing, Swinburne University of Technology, PO Box 218, Hawthorn, VIC 3122, Australia}
\affiliation{ARC Centre of Excellence for All-sky Astrophysics (CAASTRO)}

\author{Wael Farah}
\affiliation{Centre for Astrophysics and Supercomputing, Swinburne University of Technology, PO Box 218, Hawthorn, VIC 3122, Australia}

\author{Anne J. Green}
\affiliation{Sydney Institute for Astronomy (SIfA), School of Physics, The University of Sydney, NSW 2006, Australia}

\author{Vivek Gupta}
\affiliation{Centre for Astrophysics and Supercomputing, Swinburne University of Technology, PO Box 218, Hawthorn, VIC 3122, Australia}

\author[0000-0002-3205-8288]{Richard W. Hunstead}
\affiliation{Sydney Institute for Astronomy (SIfA), School of Physics, The University of Sydney, NSW 2006, Australia}

\author[0000-0002-0996-3001]{Andrew Jameson}
\affiliation{Centre for Astrophysics and Supercomputing, Swinburne University of Technology, PO Box 218, Hawthorn, VIC 3122, Australia}
\affiliation{ARC Centre of Excellence for All-sky Astrophysics (CAASTRO)}

\author{Ayushi Mandlik}
\affiliation{Centre for Astrophysics and Supercomputing, Swinburne University of Technology, PO Box 218, Hawthorn, VIC 3122, Australia}

\author[0000-0003-0289-0732]{Stefan Os{\l}owski}
\affiliation{Centre for Astrophysics and Supercomputing, Swinburne University of Technology, PO Box 218, Hawthorn, VIC 3122, Australia}

\author{Aditya Parthasarathy}
\affiliation{Centre for Astrophysics and Supercomputing, Swinburne University of Technology, PO Box 218, Hawthorn, VIC 3122, Australia}
\affiliation{CSIRO Astronomy and Space Science, Australia Telescope National Facility, Epping, NSW 1710, Australia}

\author[0000-0003-2783-1608]{Daniel C. Price}
\affiliation{Centre for Astrophysics and Supercomputing, Swinburne University of Technology, PO Box 218, Hawthorn, VIC 3122, Australia}
\affiliation{Department of Astronomy, University of California Berkeley, 501 Campbell Hall, Berkeley CA 94720}

\author{Angus Sutherland}
\affiliation{Sydney Institute for Astronomy (SIfA), School of Physics, The University of Sydney, NSW 2006, Australia}

\author{David Temby}
\affiliation{Sydney Institute for Astronomy (SIfA), School of Physics, The University of Sydney, NSW 2006, Australia}

\author{Glen Torr}
\affiliation{Sydney Institute for Astronomy (SIfA), School of Physics, The University of Sydney, NSW 2006, Australia}

\author{Glenn Urquhart}
\affiliation{Sydney Institute for Astronomy (SIfA), School of Physics, The University of Sydney, NSW 2006, Australia}

\author[0000-0001-9518-9819]{Vivek Venkatraman Krishnan}
\affiliation{Max-Plank-Institute f{\"u}r Radioastronomie, Auf dem H{\"u}gel 69, D-53121 Bonn, Germany}

\keywords{stars: neutron --- pulsars: individual (PSR J0908$-$4913)}

\section{}

Neutron star glitches can originate from either a transfer of angular momentum from the core to the crust via the unpinning of superfluid vortices~\citep{Anderson1975} or cracking of the star's crust~\citep{Baym1969, Ruderman1969}. Independent of the  underlying mechanism, glitches result in a near instantaneous increase in the observed spin frequency and are sometimes associated with a change in spin-down. Occasionally, the change in spin frequency is observed to exponentially recover toward the pre-glitch value.
Studies of large catalogues of glitches have provided insights to the internal dynamics of neutron stars and the behaviour of matter under super-nuclear densities~\citep[e.g.][]{Yu2013}.

We report the first detection of a glitch in the radio pulsar PSR J0908$-$4913 (PSR B0906$-$49)  during regular timing observations by the Molonglo Observatory Synthesis Telescope (MOST) as part of the UTMOST project~\citep{Bailes2017}. MOST is a 1.6-km long cylindrical aperture synthesis telescope with a diameter of 11.7\,m, located approximately 40\,km to the South-East of Canberra, Australia. It currently observes right-hand circularly polarised radio emission at a centre frequency of 835\,MHz with a bandwidth of 31.25\,MHz.

PSR J0908$-$4913 was initially discovered by MOST 31\,years ago~\citep{DAmico1988}. It is a bright ($35 \pm 3$\,mJy at 843\,MHz) pulsar with a spin period of $107$\,ms and dispersion measure of $180.37 \pm 0.05$\,pc\,cm$^{-3}$. It has been regularly timed by the Parkes 64-m telescope for over two decades~\citep[][Johnston, private communication]{Weltevrede2010,Yu2013} and as part of the UTMOST pulsar timing programme since May 2015~\citep[see][]{Jankowski2019}. It has not been observed to glitch throughout its entire timing history, until now.

We can parameterise the associated change in rotation phase from a glitch by measuring the permanent changes in spin-frequency ($\Delta\nu_{p}$) and spin-down frequency ($\Delta\dot{\nu}_{p}$), in addition to an exponential spin recovery ($\Delta\nu_{d}$) over some timescale ($\tau_{d}$) as
\begin{equation}
    \phi(t) = \Delta \phi + \Delta\nu_{p}(t - t_{g}) + \frac{1}{2}\Delta\dot{\nu}_{p}(t -t_{g})^{2} - \Delta\nu_{d}\tau_{d} e^{-(t - t_{g})/\tau_{d}},
\end{equation}
where $\Delta \phi$ is an unphysical jump in phase to account for ambiguities in the exact number of rotations since the glitch epoch ($t_{g}$).
The glitch we detected occurred between observations taken on MJD 58762.92 and MJD 58771.90.
After fitting over a grid of $t_{g}$ to find the date with the smallest $\Delta\phi$, we constrained the glitch epoch to MJD $58765.06 \pm 0.05$ (UTC 2019-10-09-01:26:00 $\pm$ 12\,minutes).
Performing parameter estimation with \texttt{TempoNest}~\citep{Lentati2014}, we find the glitch is best described by a permanent change in spin-frequency of $\Delta\nu_{p} = 203.6 \pm 1.2 \times 10^{-9}$\,Hz, with no evidence for a change in spin-down or spin recovery to date.\footnote{A figure showing the timing residuals can be found here: \href{https://astronomy.swin.edu.au/research/utmost/?p=1805}{astronomy.swin.edu.au/research/utmost/?p=1805}} Additional post-glitch observations will better constrain any changes in spin-down or recovery. The measured glitch parameters and upper-limits are summarised in Table~\ref{tbl:params}

With an amplitude of $\Delta\nu_{g}/\nu = 21.7 \pm 0.1 \times 10^{-9}$, this glitch is similar to those seen in pulsars with similar spin-down rates. Continued monitoring of this pulsar is being undertaken by UTMOST. Attempts to measure any long-term recovery and pulse shape changes will be the subject of future works.

\begin{deluxetable}{ll}
\tablecaption{Median posterior glitch values and $68\%$ confidence intervals.\label{tbl:params}}
\tablehead{
\colhead{Parameter} & \colhead{Value}
}
\startdata
$t_{g}$ (MJD)                        & $58765.06 \pm 0.05$             \\
$\Delta\nu_{p}$ (Hz)                 & $203.6 \pm 1.2 \times 10^{-9}$  \\
$\Delta\dot{\nu}_{p}$ (Hz\,s$^{-1}$) & $\lesssim -1.61 \times 10^{-15}$ \\
$\Delta\nu_{g}/\nu$ ($\times 10^{-9}$)   & $21.7 \pm 0.1$                  \\
\enddata
\end{deluxetable}

\acknowledgments

The Molonglo Observatory is owned and operated by the University of Sydney. Major support for the UTMOST project has been provided by Swinburne University of Technology. We acknowledge the Australian Research Council grants CE110001020 (CAASTRO) and the Laureate Fellowship FL150100148. M.E.L. acknowledges support from the Australian Government Research Training Program and CSIRO Astronomy and Space Science. This work made use of the OzSTAR national HPC facility. OzSTAR is funded by Swinburne University of Technology and the National Collaborative Research Infrastructure Strategy.

\bibliographystyle{aasjournal}
\bibliography{glitch}

\end{document}